\documentclass[a4paper,twocolumn,11pt,accepted=2022-12-07]{quantumarticle}
\pdfoutput=1
\usepackage[utf8]{inputenc}
\usepackage[english]{babel}
\usepackage[T1]{fontenc}
\usepackage{amsmath}
\usepackage{hyperref}

\usepackage{graphicx}
\usepackage{ulem}
\usepackage{gensymb}
\usepackage{bm}
\usepackage[numbers,sort&compress]{natbib}
\usepackage{dsfont}
\usepackage{epsfig}
\usepackage{feynmf}
\usepackage{blindtext, rotating}
\usepackage{mathtools}
\usepackage{dsfont}
\usepackage{subcaption}
\usepackage{physics}
\usepackage{amsfonts}
\usepackage{ragged2e}
\usepackage{siunitx}
\usepackage{comment}
\usepackage{soul}
\usepackage{lipsum}
\usepackage{hyperref} 

\usepackage{xcolor}

\begin{document}

\title{Quantum signatures in nonlinear gravitational waves}

\author{Thiago Guerreiro}
\email{barbosa@puc-rio.br}
\affiliation{Department of Physics, Pontifical Catholic University of Rio de Janeiro, Rio de Janeiro 22451-900, Brazil}
\author{Francesco Coradeschi}
\email{fr.coradeschi@gmail.com}
\affiliation{Istituto del Consiglio Nazionale delle Ricerche, OVI, Italy}
\author{Antonia Micol Frassino}
\email{antoniam.frassino@icc.ub.edu}
\affiliation{Departament de F{\'\i}sica Qu\`antica i Astrof\'{\i}sica, Institut de Ci\`encies del Cosmos,
Universitat de Barcelona, Mart\'{\i} i Franqu\`es 1, E-08028 Barcelona, Spain}
\author{Jennifer Rittenhouse West}
\email{jennifer@lbl.gov}
\affiliation{Lawrence Berkeley National Laboratory, Berkeley, CA 94720, USA}
\author{Enrico Junior Schioppa}
\email{enrico.schioppa@unisalento.it}
\affiliation{Dipartimento di Matematica e Fisica ``E. De Giorgi'', Universit\`a del Salento, and Istituto Nazionale di Fisica Nucleare (INFN) sezione di Lecce, via per Arnesano, 73100 Lecce, Italy}

\maketitle

\begin{abstract}
    The effective quantum field theory description of gravity, despite its non-renormalizability, allows for predictions beyond classical general relativity. 
    As we enter the age of gravitational wave astronomy, an important and timely question is whether measurable quantum predictions that depart from classical gravity, analogous to quantum optics effects which cannot be explained by classical electrodynamics, can be found.  In this work, we investigate quantum signatures in gravitational waves using tools from quantum optics. Squeezed-coherent gravitational waves, which can exhibit sub-Poissonian graviton statistics, can enhance or suppress the signal measured by an interferometer, a characteristic effect of quantum squeezing. Moreover, we show that Gaussian gravitational wave quantum states can be reconstructed from measurements over an ensemble of optical fields interacting with a single copy of the gravitational wave, thus opening the possibility of detecting quantum features of gravity beyond classical general relativity. 
\end{abstract}

\section{Introduction}

Gravitational wave (GW) detectors have opened a new window into the astrophysical universe, enabling the first direct observations of gravitational radiation from collisions of compact objects \cite{Nitz2021}. 
From a quantum mechanical point-of-view, interferometers such as LIGO and VIRGO act as field quadrature detectors for GWs \cite{Pang2018, Guerreiro2020}, much like homodyne detectors in quantum optics \cite{Davidovich1996}. 
It is therefore natural to ask whether GW interferometers could be used to probe the quantum mechanical nature of gravity \cite{Dyson2013} in analogy to quantum optics \cite{doi:https://doi.org/10.1002/9781119009719.ch5}. 

It is worth taking a moment to specify the meaning of \textit{quantum mechanical nature of gravity} in this context. While a complete theory of quantum gravity is at this time unrealized, there are no conceptual problems in quantizing the Einstein-Hilbert action with the understanding that it is an \textit{effective} theory, valid well below a cut-off scale assumed to be roughly of the order of the Planck scale by power counting arguments. This approach is well-suited to studying potential quantum effects of gravitational waves in the range of energies and frequencies current experiments are sensitive to. For a recent review see for instance \cite{Coradeschi2021}.

Recently, the effect of quantum mechanical noise associated to graviton fluctuations of GWs on a model detector has been calculated \cite{Parikh2021b} using the effective quantum field theory description of gravitational waves, and the associated modification of Newton's law discussed in \cite{chawla2021}.
It was pointed out that the gravity-induced fluctuations -- \textit{noise} -- in the motion of test masses associated to squeezed GW states are enhanced by an exponential factor in the squeezing parameter \cite{Guerreiro2020, Parikh2021a}, giving rise to renewed hopes of detecting quantum signatures of the gravitational field in future detectors.
These findings lead to the central questions of this work: (i) are there any quantum GW states, in the specific context of quantum Einstein-Hilbert effective field theory, that would produce exponential enhancement factors in a \textit{signal} (apart from the gravity-induced fluctuations) detectable by GW detectors and (ii) could we use the characteristics of such signals to discriminate different quantum GW states?
As we will show, both questions can be positively answered.

It is crucial to stress that our focus, at this stage, is not on the actual production of such quantum GW states in astrophysical processes, but rather on their \textit{in principle} detectability using present-day or near-future experimental setups. As it turns out, our main candidate quantum GW states, \textit{squeezed-coherent states}, do have potential production mechanisms in astrophysics. 
Squeezed GWs are expected to be generated in highly nonlinear situations and at present proposed sources are inflationary cosmology~\cite{PhysRevD.42.3413, PhysRevD.50.4807,PhysRevD.55.5917} and the final stages of black hole evaporation~\cite{natueHawking}, with GW squeezing recently bounded using LIGO observations \cite{Hertzberg2022}. 
It has also been suggested that highly nonlinear perturbed black holes might lead to the production of noncoherent quantum states of GWs; see \cite{Parikh2021b} and \cite{Parikh2021a}. 

Squeezed-coherent states, in contrast to the simpler coherent states, can be thought of as inherently quantum mechanical, exhibiting interference in phase space \cite{Wheeler1987} and the possibility of sub-Poissonian statistics \cite{Davidovich1996}.
Measuring characteristic features produced by such states would offer an empirical path to probe quantum mechanical effects related to gravity. 
As we will show, squeezed-coherent GWs with modest -- {\it i.e.} order one -- squeezing parameters can produce enhanced  or suppressed signals in an optical interferometer with sensitivity on the order of present-day GW detectors, in addition to the induced fluctuations associated to quantum uncertainty. This provides a signature of the quantum nature of the waves; the precise value of the enhancement or suppression depends on the peculiarities of the source that produces the waves and are controlled by a phase parameter associated to the state. Furthermore, measurements over an ensemble of optical fields populating the interferometer can be used to completely reconstruct Gaussian quantum states of GWs, opening the possibility of witnessing non-classical features of gravity, such as sub-Poissonian graviton number statistics.
Note that contrary to their optical counterparts, squeezed-coherent states of gravitons are expected to propagate through the universe nearly unperturbed \cite{MTW}, carrying information about their sources with preserved coherence.

Overall, our findings integrate part of an emerging subfield of fundamental physics in which ideas from quantum optics offer new paths to uncover unexplored natural phenomena, with other key examples including the search for beyond the standard model physics using atomic and molecular physics \cite{Safronova2018}, fifth forces with quantum optomechanics \cite{Monteiro2020, Blakemore2021, Moore2021}, quantum-enhanced searches for light dark matter \cite{Backes2021, Aybas2021, Graham2018, Wurtz2021}, and nonlinear optics and optomechanics \cite{Estrada2021, Carney2021}.
Quantum gravity itself is not exempt. Of particular interest to us are experimental proposals seeking to measure quantum effects associated to gravity, as well as to measure the gravitational field of quantum matter in table-top experiments \cite{Krisnanda2017, Bose2017, Marletto2017, Oniga2016, Carney2021b, Carney2021c, Streltsov2022, Westphal2021}. While very important, these proposals rely on challenging quantum engineering \cite{ Aspelmeyer:2022fgc}, whereas we consider an alternative observational approach, of seeking for natural phenomena in which quantum effects of gravity are manifest.

This paper is organized as follows: in Sec.~\ref{sec:squez-cohe}
we briefly describe a model interferometric GW detector and its interaction with canonically quantized weak gravitational field perturbations. In Sec.~\ref{sec:GW-qubit} we describe the interaction between GW states and a simplified toy model detector consisting of an ensemble of optical qubits. Sec.~\ref{subpoisson} is devoted to squeezed-coherent GWs  and in Sec. \ref{reconstruction} we describe how Gaussian GW quantum states can be reconstructed from optical measurements. Sec. \ref{Squezed-coherent-GWs} describes the interaction between optical coherent states (such as the laser field of a GW interferometer) and squeezed-coherent GWs. Finally, in Sec.~\ref{sec:conclusions} we discuss future perspectives on the generation and detection of nonlinear quantum GWs. For details on the calculations throughout the text, we refer to the Appendices.

\section{Mode-detector interaction}
\label{sec:squez-cohe}

We consider the interaction of a weak GW with a single mode cavity electromagnetic field as a model for our interferometric GW detector. The model captures the essential features of the effects we aim to discuss. Detectors like LIGO, a variant of the Michelson interferometer, can be formally mapped to a Fabry-P\`erot cavity \cite{Jarzyna2015}. 
The treatment in this work is the same as in \cite{Pang2018, Oniga2016, Bose2017, Toros2020, Buonanno2003} and \cite{Guerreiro2020} for the canonical quantization of weak metric perturbations around a flat background and we refer to those works for further details on the formalism while we briefly summarize the main results necessary for our discussion.

The interaction between a cavity electromagnetic field and quantized metric perturbations in the tranverse traceless gauge can be formally derived from the Einstein-Hilbert action coupled to the Maxwell stress-energy tensor (see \cite{Pang2018} for more details). 
For a single mode electromagnetic cavity with photon annihilation (creation) operator $ a $ ($ a^{\dagger} $) in the presence of $ + $ polarized GWs propagating along the $ \bm{n} = \hat{z} $ direction, perpendicular to the cavity axis (chosen to be along the $ \hat{x} $ direction), the interaction reads
\begin{eqnarray}
H_{I} = - \dfrac{\omega}{4} a^{\dagger} a \int \dfrac{d \bm{k}}{\sqrt{(2\pi)^{3}}} \left(   \sqrt{\dfrac{8\pi G}{k}} \mathfrak{b}_{\bm{k}} + h.c.  \right)
\label{H_int_canonical}
\end{eqnarray}  
where $ k = \vert \bm{k} \vert = \Omega_{k} $ is the GW frequency for the mode $ \bm{k} $, the operator $  \mathfrak{b}_{k}^{\lambda} $ ($ \mathfrak{b}_{\bm{k}}^{\lambda \dagger} $) is the canonical graviton annihilation (creation) operator and $ \omega $ is the cavity frequency. \\
For our purposes, it is convenient to introduce a discrete form of \eqref{H_int_canonical}. 
Following the standard procedure in quantum optics \cite{Scully1997} we introduce a quantization volume $ V $ and note that $ \left[ \sqrt{8\pi G / k} \right] = L^{3/2} $, where $ L$ denotes the dimension of length. Further, $ \left[ d \bm{k} \right] = L^{-3} $, which implies the graviton annihilation and creation operators have dimension $ \left[ \mathfrak{b}_{\bm{k}} \right] = L^{3/2} $. Define $ \mathfrak{b}_{\bm{k}} = \sqrt{V}  b_{\bm{k}} $, where $  b_{\bm{k}} $ is a dimensionless quantity.  The discrete limit $ d \bm{k} \rightarrow 1 / V $, $ (2\pi)^{-3/2}\int \rightarrow \sum $ then leads to 
\begin{eqnarray}\label{eq:H_I}
H_{I} = - \dfrac{\omega}{4} a^{\dagger} a \sum_{\bm{k}} \left(   \sqrt{\dfrac{8\pi G}{kV}} b_{\bm{k}} + h.c. \right).
\end{eqnarray}
Defining the single graviton strain for the mode $ \bm{k} $ as $ f_{\bm{k}} =  \sqrt{8\pi G/(kV)} $, then the coupling constant of a GW mode $ \bm{k} $ with the cavity electromagnetic field is $ g_{\bm{k}} = \omega  f_{\bm{k}} / 4 $, which has dimension of frequency. It is also convenient to introduce the dimensionless coupling $ q_{\bm{k}} =  g_{\bm{k}} / \Omega_{k} $.
Therefore, taking into account the previous interaction term, the total discretized Hamiltonian reads $H = H_{0} + H_{I}$, with $H_0$ the free-field Hamiltonian for the cavity and GW modes: 
\begin{eqnarray}\label{eq:H_0}
H_{0} = \omega a^{\dagger} a + \sum_{\bm{k}} \Omega_{k} b_{\bm{k}}^{\dagger} b_{\bm{k}} \,.
\end{eqnarray}

The time evolution generated by the total Hamiltonian 
can be calculated exactly  and is given by 
\begin{eqnarray}
U(t) = e^{-i H t} = \prod_{\bm{k}} U_{\bm{k}}(t)\,,
\label{total_U}
\end{eqnarray}
where the unitary operator in the interaction picture reads:
\begin{eqnarray}
U_{\bm{k}} (t)  = e^{iB_{k}(t) (a^{\dagger}a)^{2}} e^{q_{\bm{k}} a^{\dagger} a  (\gamma_{k} b^{\dagger}_{k} -\gamma_{k}^{*} b_{k})} \,,
\label{Uk}
\end{eqnarray}
with $ \gamma_{k} = (1 - e^{-i\Omega_{k} t}) $ and $B_{k}(t) = q_{\bm{k}}^{2} \left(   \Omega_{k} t - \sin \Omega_{k} t  \right)   $  and the accompanying GW states evolve according to the free evolution \cite{Brandao2020, Guerreiro2020}
\begin{eqnarray}
\vert \Psi(t) \rangle = \prod_{\bm{k}} e^{-i b_{\bm{k}}^{\dagger} b_{\bm{k}} \Omega t} \vert \Psi \rangle \,.
\end{eqnarray}
See Appendix \ref{append:time} for further details.
Note that the $ B_{k}(t) $ term is quadratic in $ q_{\bm{k}} $, which leads to effects of order $ G$ in the exponent. As shown in \cite{Guerreiro2020}, this term results in sub-leading effects for various states of interest, such as the vacuum, coherent, squeezed vacuum and thermal states of GWs. From now on, we will therefore approximate \eqref{Uk} as
\begin{eqnarray}\label{eq:U(t)approx}
U_{\bm{k}} (t)  \approx e^{q_{\bm{k}} a^{\dagger} a  (\gamma_{k} b^{\dagger}_{k} -\gamma_{k}^{*} b_{k}    )}\,,
\label{Uk_aprox}
\end{eqnarray}
where $\gamma_k$ 
incorporates the time dependence. 
Physically, the meaning of \eqref{Uk_aprox} is clear: it corresponds to a GW-induced \textit{phase operator} acting on the electromagnetic mode, with the acquired phase proportional to the GW quadrature. Conversely, \eqref{Uk_aprox} acts on the GW mode as a \textit{displacement} operator, proportional to the number of photons contained in the electromagnetic field.

We are interested in the dynamics of an optical field interacting with a single GW mode, but notice the unitary operator \eqref{total_U} couples the detector to \textit{all} GW modes. As it turns out \cite{Guerreiro2020}, the effect of additional modes populated by the vacuum or thermal states at the expected temperature for a cosmic GW background yields negligible corrections to the dynamics of a single-mode GW populated by a macroscopic mean number of gravitons, such as the ones detectable by LIGO. 
Therefore, from now on we will make a single-mode approximation, but for estimates on the corrections due to the additional `empty' modes see Appendix \ref{multi-mode-corrections}.

\section{Interaction between a quantum GW and a qubit}
\label{sec:GW-qubit}

With the operator \eqref{Uk_aprox} we can calculate the time evolution of arbitrary states of the optical field interacting with the relevant modes of quantum GWs. 
To illustrate that, consider the simplified example of an optical qubit spanned by the vacuum and single photon states $ \left\{ \vert 0 \rangle, \vert 1 \rangle \right\}$.
For simplicity, we drop the index $\bm{k}$ and unless specified consider only a single GW mode.
Time evolution of the optical qubit in the presence of a GW in the state $ \vert \Psi(t) \rangle $ is then
\begin{align}
\left( \alpha \vert 0 \rangle  +  \beta \vert 1 \rangle \right) \vert \Psi(t) \rangle &\rightarrow \nonumber
\\
 \alpha \vert 0 \rangle \vert \Psi(t) \rangle &+ \beta \vert 1 \rangle D(q\gamma) \vert \Psi(t) \rangle \, ,
\end{align}
where $D(q\gamma)$ is the displacement operator acting on the GW mode and $ q, \gamma $ are the dimensionless coupling and time-evolution function as defined in the previous section, specialized for the mode of interest.
Notice the states $  \vert \Psi(t) \rangle $ and $ D(q\gamma) \vert \Psi(t) \rangle $ are in general not orthogonal.
Tracing out the GW mode, we find the time-dependent qubit density matrix,
\begin{equation}\label{eq:densityMatrixTracedOut}
    \rho \left(t\right) = \begin{pmatrix}
    \vert \alpha \vert^{2} & \mathcal{I}_{1}(t) \alpha \beta^{*} \\
    \mathcal{I}_{1}(t)^{*} \alpha^{*} \beta & \vert \beta \vert^{2}
    \end{pmatrix}
\end{equation}
where,
\begin{eqnarray}
\mathcal{I}_{1}(t) = \langle \Psi(t) \vert D(q\gamma) \vert \Psi(t) \rangle\,.
\label{Bloch2}
\end{eqnarray}
Dynamics of the qubit in the Bloch sphere is governed by the inner product $ \mathcal{I}_{1}(t) $, which is a complex number with $\vert \mathcal{I}_{1}(t) \vert \leq 1 $. 

\begin{figure}[t!]
    \centering
    \includegraphics[width=0.4\textwidth]{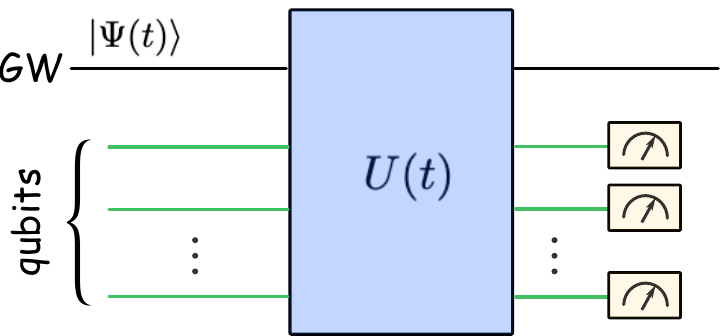}
    \caption{Conceptual experiment of a quantum GW interacting with a simplified toy model detector consisting of an ensemble of optical qubits. The GW state induces an open quantum system dynamics on the qubits, and noise spectroscopy can in principle be used to discriminate between different quantum GW states and measure quantum properties of the gravitational field.} 
    \label{concpet}
\end{figure}

The modulus of $ \mathcal{I}_{1}(t) $ is associated to GW induced decoherence processes: it multiplies the off-diagonal terms of the density matrix, and as such it increases the von Neumann entropy of the optical mode whenever $ \vert \mathcal{I}_{1}(t) \vert < 1 $. 
An increase in entropy can be understood as a consequence of noise in the system.
For certain situations, such as the case of a multi-mode GW thermal state at temperature $ T $, Markovian approximations can be made and it is possible to show that interaction of the qubit with the GW field leads to a decoherence rate $\Gamma $ proportional to $ k_{B} T \left( \Delta E / E_{\rm pl} \right)^{2} $, where $ \Delta E $ is the energy difference between qubit states and $ E_{\rm pl} $ is the Planck energy \cite{Blencowe2013}; see Appendix \ref{appendix-2-decoherence} for a derivation. 

In general the time evolution comprises a combination of shrinkage of the Bloch vector, recoherence, and a unitary evolution given by the phase of $ \mathcal{I}_{1}(t) $ resulting in precession of the Bloch vector. 
Casting the qubit density matrix as a linear combination of Pauli matrices $\sigma_i$ in the standard form 
\begin{equation}
\rho(t) = \frac{\mathds{1} + \vec{r}(t)\cdot \vec{\sigma}}{2}\,,
\end{equation}
we can explicitly calculate the Bloch vector $\vec{r}(t)$ for GWs in different quantum states.
Considering as the qubit initial state $\rho(0) = \vert + \rangle \langle + \vert$, the Bloch vectors of qubits interacting with GWs in the vacuum $ ( \vert 0 \rangle ) $, coherent $( \vert h \rangle ) $ and squeezed $( \vert r \rangle ) $ states are respectively given by,
\begin{align}
&\vec{r}_{\vert 0 \rangle}(t) = \  e^{-q^{2} \left( 1 - \cos \Omega t \right)} \left( \cos \omega t, \sin \omega t, 0  \right),  \label{Bloch_vec1} \\
&\vec{r}_{\vert h \rangle}(t) = \  e^{-q^{2} \left( 1 - \cos \Omega t \right)} \times \nonumber \\
& \left( \cos \left( \omega t + \Delta \varphi_{0}(t) \right),  \sin \left( \omega t + \Delta \varphi_{0}(t) \right), 0  \right), \label{Bloch_vec2} \\
&\vec{r}_{\vert r \rangle}(t) = \  e^{-8 e^{2r} q^{2} \sin^{4}\left(\frac{\Omega t}{2} \right)} \left( \cos \omega t, \sin \omega t, 0  \right),  \label{Bloch_vec3} 
\end{align}
where the phase $ \Delta \varphi_{0}(t) $ in the coherent case is (see the discussion in the next sections),
\begin{eqnarray}
\Delta \varphi_{0}(t) = 2 q \vert h \vert \sin \Omega t \ .
\end{eqnarray}
In deriving \eqref{Bloch_vec1}, \eqref{Bloch_vec2} and \eqref{Bloch_vec3} we have considered a squeezed state with vanishing squeezing angle and have taken the limit $ r \gg 1 $, for simplicity, but generalization to arbitrary values of $ r $ and squeezing phase is straightforward.
The coupling strength $ q $ governing the interaction between the optical and GW fields is small \cite{Guerreiro2020}, which guarantees that the backreaction of the detector upon the gravitational field can be neglected \cite{Pang2018}.
The prefactors in the Bloch vectors account for periodic shrinking and as discussed are associated to gravitational induced \textit{noise}, while the phase $ \Delta \varphi_{0}(t) $ in the coherent case causes a periodic precession of the qubit state analogous to the phase signal imprinted upon a coherent optical mode by a GW. As we will soon show, the phase $ \Delta \varphi_{0}(t) $ is also present in the case of a GW interacting with a coherent electromagnetic state and corresponds to the signal measured by interferometric GW detectors. 
Notice also the appearance of the exponential enhancement factor $e^{2r} $ in the shrinking factor associated to squeezed GW states, as predicted in \cite{Guerreiro2020, Parikh2021a, Parikh2021b}. 

Trajectories of the Bloch vector are hence dependent on the quantum GW state. 
Different GW quantum states can \textit{in principle} be discriminated by qubit noise spectroscopy \cite{Clerk2010}. 
One could then conceive a conceptual experiment as shown in Figure \ref{concpet}, in which an ensemble of qubits is left to interact with a GW state for some time, after which measurements on the qubits are performed. 
Properties of the statistical distribution of measured states will then contain information on the GW quantum state. This scheme offers a path to search for quantum gravitational effects using interferometric detectors and to discriminate between different quantum states of the GW field.
The conceptual experiment shown in Figure \ref{concpet} can be understood as a quantum mechanical version of the proposals in \cite{Parikh2021a, Parikh2021b} where, in the context of a classical detector consisting of two test masses separated by a certain distance, it was shown that fluctuations associated to GWs in different quantum states lead to modifications of the geodesic deviation equation by addition of a stochastic Langevin term whose noise correlators are specific to the quantum GW state. Observation of the stochastic trajectories of the test masses could then reveal quantum features of the GW field. 
Analogously, observation of the stochastic \textit{quantum} trajectories of an ensemble of optical states as dictated by \eqref{Bloch_vec1} - \eqref{Bloch_vec3} would allow for the discrimination of different quantum states of GWs.
Notably, the authors of \cite{Parikh2021a, Parikh2021b} have also shown that when the GW is in a squeezed state with squeezing parameter $ r $, statistical properties of the trajectories of test masses such as the two-point position correlation function and consequently the position power spectral density depend on an enhancement factor approximatelly given by $e^{2r}$, as we also found in the Bloch vector \eqref{Bloch_vec3}.

\section{Squeezed-coherent GW states}
\label{subpoisson}
We have considered thus far coherent and squeezed states of the GW field. We turn our attention to single-mode \textit{squeezed-coherent states}, defined as 
\begin{eqnarray}
\vert \xi, h \rangle = D(h) S(\xi) \vert 0 \rangle \, , 
\label{DS}
\end{eqnarray}
where $ S(\xi) $ is the squeezing operator, $ h = \vert h \vert e^{i\theta} $ and $ \xi = r e^{i\phi} $. Note that the displacement and squeezing operators do not commute but we can convert squeezed-displaced waves ($ SD $) into displaced-squeezed states ($ DS  $) via the operator identity,
\begin{eqnarray}
D(h)S(\xi) = S(\xi) D(h \cosh r + h^{*} e^{2i\phi} \sinh r) \, . \nonumber \\
\label{SD}
\end{eqnarray}
In other words, one form can always be brought into the other through  a suitable redefinition of constants. 
We denote gravitational coherent states by $ \vert h \rangle = D(h) \vert 0 \rangle $, see Appendix \ref{append:time} for definitions and useful formulae. 

Squeezed-coherent states can be non-classical in the sense that their graviton number statistics can become sub-Poissonian. To see that, define the Fano factor,
\begin{equation*}
    F = \Delta N^{2} / \langle N \rangle \, ,
\end{equation*}
where $ \Delta N^{2}, \langle N \rangle $ are the variance and mean number of gravitons, respectively.
Classical states are obtained by mixing coherent states and always have $ F \geq 1 $ \cite{Scully1997} -- i.e. classical states are super-Poissonian -- while a Fano factor smaller than one -- sub-Poissonian statistics -- can only be explained by quantum theory.
The Fano factor is related to the field second-order autocorrelation function $ g^{2}_{GW} = \langle b^{\dagger} b^{\dagger} b b \rangle / \langle b^{\dagger} b \rangle^{2} $ as 
\begin{eqnarray}
g^{2}_{GW} = 1 + \left( F - 1 \right)/\langle N \rangle \, ,
\end{eqnarray}
so $ F < 1 $ implies $ g^{2}_{GW} < 1$. 
For a squeezed-coherent state as defined in \eqref{DS} with $ \phi = 2 \theta $, the variance and mean graviton numbers are given by \cite{Davidovich1996},
\begin{eqnarray}
\Delta N^{2} &=&  \vert h \vert^{2} e^{-2r} + 2 \sinh^{2} r \cosh^{2} r \, ,
\\
\langle N \rangle &=& \vert h \vert^{2} + \sinh^{2} r \, .
\end{eqnarray}
In the limit that $ \vert h \vert^{2} \gg 2 e^{2r} \sinh^{2} r \cosh^{2} r  $ we have
\begin{eqnarray}
F = \dfrac{\Delta N^{2}}{\langle N \rangle} \approx e^{-2r} < 1 \, .
\end{eqnarray}
Note that for a squeezing of order $ r \approx 1 $, the above approximation is valid whenever $ \vert h \vert^{2} \gg 50 $. 
In this sense squeezed-coherent states can display non-classical behavior even in the macroscopic limit of large numbers of quanta. While this is also the case in electromagnetic quantum optics \cite{oudot2015two}, photons strongly interact with matter, so these states tend to be short-lived \cite{zurek1993coherent}. In contrast, GWs interact very weakly with their environment. If squeezed-coherent GW states are produced in nature, they could travel the universe nearly unperturbed \cite{Misner2009}.

\section{GW state reconstruction and non-classical observables}
\label{reconstruction}

We would now like to show how quantum information contained in a GW can be obtained from an experiment as depicted in Figure \ref{concpet}, where a single copy of the GW interacts with multiple copies of an optical state and subsequent measurements are performed on the optical ensemble. As it turns out, if the GW state is Gaussian -- as is the case for squeezed-coherent GWs -- we could in principle completely reconstruct the gravitational quantum state and consequently perform non-classicality witness measurements and also obtain the gravitational second-order coherence $ g^{2}_{GW} $.

Consider a single mode GW initially in the state $ \vert \Psi \rangle $ interacting with the optical wavefunction,
\begin{eqnarray}
\vert \Phi \rangle = a_{0} \vert 0 \rangle + a_{1} \vert 1 \rangle + a_{2} \vert 2 \rangle + ...
\label{optical_state}
\end{eqnarray}
The total initial state is $ \vert \Phi \rangle \vert \Psi \rangle  $. Time evolution according to \eqref{Uk_aprox} reads,
\begin{eqnarray}
U \vert \Phi \rangle \vert \Psi \rangle =   a_{0} \vert 0 \rangle \vert \Psi \rangle + a_{1} \vert 1 \rangle  D(-q\gamma) \vert \Psi \rangle   \nonumber \\
+ a_{2} \vert 2 \rangle D(-2q\gamma) \vert \Psi \rangle+ ...
\end{eqnarray}
where we have absorbed the free dynamics of the GW in the displacement operator, $ \langle \Psi(t) \vert D(q\gamma) \vert \Psi(t) \rangle = \langle \Psi \vert D(-q\gamma^{*}) \vert \Psi \rangle$.

The reduced density matrix of the optical mode carries information on the initial GW state. 
Denote the reduced optical density matrix components in the Fock basis as $ \rho_{nm}(t) = \langle n \vert \rho(t) \vert m \rangle  $. The density matrix is an observable and we can have access to all of its components -- say by state tomography -- provided we let several copies of the initial optical state interact with the GW. Here, we will be interested in the $ \rho_{0n} = \langle 0 \vert \rho \vert n \rangle  $ components, given by
\begin{eqnarray}
\rho_{0n} = a_{0}a_{n}^{*}  \ \langle \Psi \vert D(-nq\gamma^{*}) \vert \Psi \rangle \, ,
\end{eqnarray}
where the $ \lbrace a_{m} \rbrace $ coefficients are known, determined by the initial optical state $ \vert \Phi \rangle $.
Therefore, we have access to the quantities $ \langle \Psi \vert D(-nq\gamma^{*}) \vert \Psi \rangle $ for different values of $ n $.
Define,
\begin{align}
\mathcal{I}_{n}(t) = \langle \Psi \vert D(-qn\gamma^{*}) \vert \Psi \rangle = \nonumber \\  \mathrm{Tr}\left(  \sigma  D(-nq\gamma^{*}) \right) =  \mathrm{Tr}\left(  \sigma  e^{-qn\left(  \gamma^{*} b^{\dagger} - \gamma b  \right)} \right) \, ,
\label{I_quantity_mt}
\end{align}
where $ \sigma = \vert \Psi \rangle \langle \Psi \vert $ is the initial GW density matrix and $ \gamma^{*} = 1 - e^{it'} $, with $ t' $ denoting the rescaled dimensionless time parameter $ t' = \Omega t $. 

Recall that the \textit{quantum characteristic function} $ \chi(\eta, \eta^{*}) $ is defined as
\begin{eqnarray}
\chi(\eta, \eta^{*}) = \mathrm{Tr}\left(  \sigma e^{\eta b^{\dagger} - \eta^{*}b}  \right) = C(\eta, \eta^{*}) e^{-\vert \eta \vert^{2}/2} \, , \nonumber \\
\end{eqnarray}
where $ \eta, \eta^{*} $ are complex variables and $ C(\eta, \eta^{*}) $ is the \textit{normally ordered} quantum characteristic function,
\begin{eqnarray}
C(\eta, \eta^{*}) = \mathrm{Tr} \left( \sigma   e^{\eta b^{\dagger}}  e^{- \eta^{*} b}  \right) \, .
\label{Ncharacteristic}
\end{eqnarray}
This function contains all the information of a quantum state $ \sigma $ \cite{walls1994}, and thus can be seen as an alternative representation to the density matrix. 
It is related to the graviton-number generating function \cite{Rockower1988}, from which all normally ordered moments of the graviton number operator can be obtained, in particular the graviton second-order correlation function $ g^2_{GW} $. If we can partially reconstruct $ C(\eta, \eta^{*}) $ from data, we could obtain information on the graviton-number statistics of the corresponding GW state, or use measured moments of the GW field quadratures to witness nonclassicality of the gravitational field. 

Observe that \eqref{I_quantity_mt} corresponds to the quantum characteristic function evaluated at the contours in the complex plane given by $ qn\left( e^{it'} - 1  \right) $, so we expect that some information on the initial GW quantum state can be retrieved from measurements of the reduced optical density matrix. 
Since we only have partial access to $ C(\eta, \eta^{*}) $ in the complex domain, it is not clear how much of the information can be retrieved. It turns out that if the GW state is Gaussian, the initial GW state and consequently all of its expectation values can be completely recovered from measurements of $ \mathcal{I}_{n}(t)  $.

The essential idea is to consider time derivatives of $ \mathcal{I}_{n}(t) $ evaluated at $ t' = 0 $. For different values of $ n $, the time derivatives provide relations between normally ordered expectation values, from which we can reconstruct the desired first and second-order field correlation functions necessary for specifying the Gaussian state.  
Define the time derivatives as
\begin{eqnarray}
\alpha_{\nu}(n) \equiv \dfrac{\partial^{\nu}\mathcal{I}_{n}(t')}{\partial t'^{\nu}} \vert_{t' = 0}
\end{eqnarray}
and the field quadratures,
\begin{eqnarray}
X &=& b + b^{\dagger} \, , \label{X} \\
Y &=& i \left( b - b^{\dagger} \right) \, . \label{Y}
\end{eqnarray}
Direct computation shows that the moments are given by,
\begin{widetext}
  \begin{eqnarray}
\langle qX \rangle &=& - i \alpha_{1}(1)  \label{discrete_moments1} \, , \\
i \langle qY \rangle &=& \frac{1}{2} \alpha_{2}(2) - 2\alpha_{2}(1) \, ,  \\
\langle q^{2}X^{2} \rangle  &=& \alpha_{2}(1) - \frac{1}{2} \alpha_{2}(2) \, , \\
\langle qX qY \rangle - \frac{1}{2} \langle \left[ qX, qY \right] \rangle &=& \frac{2}{3} \alpha_{3}(1) - \frac{1}{12} \alpha_{3}(2) + \frac{1}{2}\alpha_{1}(1) \, , \\
\langle Y^{2} \rangle &=& 2 + 4\langle N \rangle - \langle X^{2} \rangle \, ,
\label{discrete_moments2}
\end{eqnarray}
\end{widetext}
where $ \langle N \rangle $ is the mean graviton number, which can be obtained by measuring the amplitude of the GW. Alternatively, $ \langle Y^{2} \rangle $ can be obtained from the other moments by normalization of the corresponding Wigner function associated to the covariance matrix. 
We recall that the time derivatives are with respect to rescaled time, and the commutator $ \left[ qX,qY  \right] = -2q^{2} i $ is known. We refer to Appendix \ref{append:reconstruction} for more details on this calculation.

From the moments $ \langle X \rangle, \langle Y \rangle, \langle X^{2} \rangle, \langle XY \rangle, \langle Y^{2} \rangle $ we can reconstruct the GW covariance matrix, which for Gaussian states is equivalent to knowledge of the full quantum state \cite{Weedbrook2012}. Furthermore, photon-number statistics, in particular the $ g^2_{GW} $ function, can be determined from knowledge of the covariance matrix \cite{Dodonov1994}.
The intracavity optical states can be determined by the leaking cavity fields via the input-output formalism \cite{vanner2015towards}.

All correlation functions above are multiplied by powers of the dimensionless coupling $ q $. We can express the moments \eqref{discrete_moments1}-\eqref{discrete_moments2} in the continuum limit by defining free field quadratures for a fixed GW polarization as
\begin{eqnarray}
h &=& \int \dfrac{dk}{\sqrt{(2\pi)^{3}}} \left( \sqrt{\dfrac{8\pi G}{k}} \mathfrak{b}_{k} +   h.c. \right) \, , \\
\bar{h} &=& i \int \dfrac{dk}{\sqrt{(2\pi)^{3}}} \left( \sqrt{\dfrac{8\pi G}{k}} \mathfrak{b}_{k} -  h.c. \right) \, .
\end{eqnarray}
Note that in retrieving the continuum limit, we write $ h, \bar{h} $ as integrals over all modes, including the empty modes discussed in Appendix \ref{multi-mode-corrections}.
We have the limits,
\begin{eqnarray}
qX \rightarrow  (\omega / 4\Omega) h \, , \\
qY \rightarrow (\omega / 4\Omega) \bar{h} \, ,
\end{eqnarray}
and
\begin{eqnarray}
\langle \left[ qX, qY \right] \rangle \rightarrow (\omega / 4\Omega)^{2} \langle \left[ h, \bar{h} \right] \rangle = -2i (\omega / 4\Omega)^{2} \, . \nonumber \\
\end{eqnarray}
Altogether, for a single-mode GW state interacting with an optical density matrix initially in state $ \rho(0) = \vert \Phi \rangle \langle \Phi \vert $ we find the relations:
\begin{widetext}
\begin{eqnarray}
\langle h \rangle &=& - \left(  \frac{4i}{\omega} \right) \frac{\partial \mathcal{I}_{1}}{\partial t} \, , \\
 \langle \bar{h} \rangle &=& \left( \frac{4i}{\omega \Omega}  \right) \left(  \frac{1}{2}\frac{\partial^{2} \mathcal{I}_{2}}{\partial t^{2}} - 2 \frac{\partial^{2} \mathcal{I}_{1}}{\partial t^{2}}  \right) \, , \\
\langle h^{2} \rangle &=& \left(  \frac{16}{\omega^{2}}  \right)  \left(   \frac{\partial^{2} \mathcal{I}_{1}}{\partial t^{2}}  - \frac{1}{2}  \frac{\partial^{2} \mathcal{I}_{2}}{\partial t^{2}}      \right)     \, , \\
\langle h\bar{h}  \rangle &=& \left(  \frac{16}{\omega^{2} \Omega}  \right) \left(  \frac{2}{3}  \frac{\partial^{3} \mathcal{I}_{1}}{\partial t^{3}}    - \frac{1}{12}  \frac{\partial^{3} \mathcal{I}_{2}}{\partial t^{3}}  \right)  + \left(  \frac{8\Omega}{\omega^{2}} \right) \frac{\partial \mathcal{I}_{1}}{\partial t} - i  \, ,  \\
\langle \bar{h}^{2} \rangle &=& 2 + 4\langle N \rangle - \langle h^{2} \rangle \, ,
\end{eqnarray}
\end{widetext}
where $ \mathcal{I}_{n}(t) =  \rho_{0n} / a_{0}a_{n}^{*} $ and the proportionality constants $ a_{0}, a_{n}^{*}  $ are determined by the initial optical state $ \vert \Phi \rangle $. Using these relations, we can obtain the quantum correlation functions of the spacetime metric perturbations associated to a single GW in the transverse traceless gauge, directly from experimentally accessible data.

\section{Optical coherent states}
\label{Squezed-coherent-GWs}

Consider now the interaction between an optical coherent state and a squeezed-coherent GW.
Let the transverse cavity electric field be defined as $ \vec{\mathcal{E}} = \mathcal{E} \hat{z} $, where $ \mathcal{E} \equiv \sqrt{\frac{\omega}{V_{c}}} \left(  a + a^{\dagger} \right) / \sqrt{2} $, with $ V_{c} $ the cavity mode volume. 
The mean electric field associated to the optical coherent state $ \vert \alpha \rangle $ interacting with an initially independent pure GW state $ \vert \Psi(t) \rangle $ reads
\begin{eqnarray}
\langle \mathcal{E}(t) \rangle = \sqrt{\dfrac{\omega}{V_{c}}} \left( \dfrac{ \mathcal{I}_{1}(t) \alpha + \mathcal{I}_{1}(t)^{*} \alpha^{*}   }{\sqrt{2}} \right)\,,
\label{field_mt}
\end{eqnarray}
where $ \mathcal{I}_{1}(t) $ is given in \eqref{Bloch2}.

For calculation of the expectation value \eqref{field_mt}, it is more convenient to write the squeezed-coherent state as a displacement followed by the squeezing operator. By use of \eqref{SD} we then write the squeezed-coherent state as,
\begin{eqnarray}
\vert \Psi(t) \rangle = S(r,\phi(t)) D(h(t)) \vert 0 \rangle\,,
\label{old_squeezed_coherent}
\end{eqnarray}
where $ S(r,\phi(t))  $ is the single-mode squeezing operator in the interaction picture with squeezing parameter $ r $ and angle $ \phi(t) = \phi_{0} - \Omega t $, and $ D(h(t)) $ is a displacement operator with amplitude $ h(t) = \vert h \vert e^{-i\Omega t}$. 

\begin{figure}[t!]
    \centering
    \includegraphics[width=0.5\textwidth]{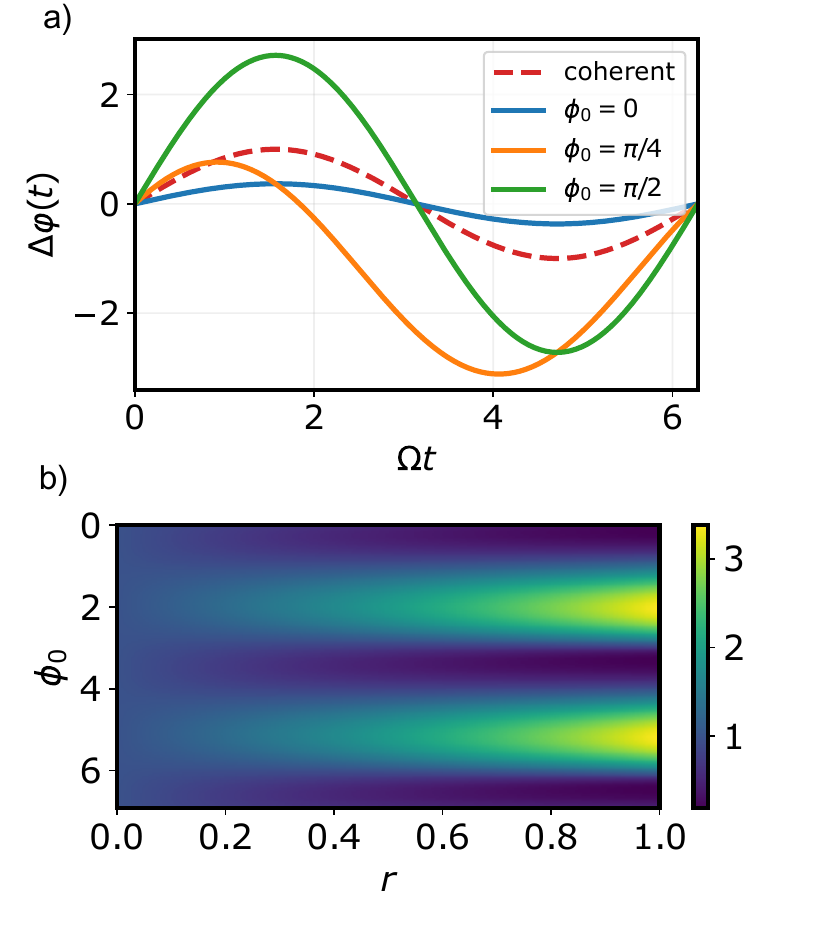}
    \caption{a) Normalized relative phase of the optical field as a function of time, in units of GW strain $ \omega f / \Omega $. Time is displayed in units of the GW period $ 2\pi / \Omega $. Different values of the squeezing angle $ \phi_{0}$ are considered, exhibiting enhancement ($\phi_{0} = \pi/2$, green curve), suppression ($\phi_{0} = 0$, blue curve) and an intermediate case ($\phi_{0} = \pi/4$, orange curve). Note that in the intermediate case the phase profile is asymmetric and exhibits a characteristic form which can be used to single out squeezed-coherent states from plain coherent states (red dashed curve).
    For this plot the squeezing parameter is taken to be $ r = 1$. b) Maximum phase value (in units of $ \omega f / 4\Omega $) as a function of squeezing parameter $ r $ and angle $\phi_{0} $ showing enhancement or suppression of the GW signal, depending on characteristics of the source. } 
    \label{phases}
\end{figure}
The $ \mathcal{I}_{1}(t) $ factor can be rewritten,
\begin{align}
& \langle \Psi(t) \vert D(q\gamma) \vert \Psi(t) \rangle = \nonumber\\
&= \langle h \vert S^{\dagger}(r,\phi) D(q\gamma) S(r,\phi) \vert  h \rangle \nonumber\\
&=  \langle h \vert D(\kappa)  \vert  h \rangle \ ,
\label{coherent_overlap}
\end{align}
where $\kappa$ denotes
\begin{eqnarray}
\kappa = q \left(  \gamma \cosh r + \gamma^{*} e^{2i\phi_{0}} e^{-2i\Omega t} \sinh r   \right)\,.
\label{kappa}
\end{eqnarray}
The above expression can be decomposed in terms of linear and quadratic contributions in $ q$. As in the qubit case, quadratic contributions yield real-valued factors and are associated with gravity-induced decoherence, which we neglect in view of the discussions in Appendix \ref{multi-mode-corrections}.
Considering only terms linear in $ q $ the relevant contribution coming from \eqref{coherent_overlap} is
\begin{eqnarray}
\exp \left( \kappa \vert h \vert e^{i\Omega t} - \kappa^{*} \vert h \vert e^{-i\Omega t} \right)\,.
\label{phase}
\end{eqnarray}
This term yields a time-varying phase $ e^{i\Delta \varphi} $ in the mean coherent electric field, dependent on the squeezing parameter $ r $ and angle $ \phi_{0} $,
\begin{align}
\Delta \varphi &= 2  \vert h \vert q \big( \sin \Omega t \cosh r +  \big. \nonumber \\
& +\big. \sin\left( 2\phi_{0} - \Omega t \right) \sinh r - \sin 2\phi_{0} \sinh r   \big)\,.
\label{general_phase}
\end{align}
This phase factor can be thought of as a continuous-variable version of the precession of the Bloch vector discussed previously, and corresponds to the phase signal measured by interferometric GW detectors.

The squeezing phase $ \phi_{0} $ is a property of the squeezed state. In electromagnetic quantum optics, for example, it is controlled by the phase of the pump beam driving the optical parametric oscillator \cite{Boyd2008}. The phase \eqref{general_phase} tells us that depending on the characteristics of a source of squeezed-coherent GWs, the signal in a GW detector can be \textit{enhanced} or \textit{suppressed} depending on $ \phi_{0} $. 
In particular, if $ \phi_{0}  = \pi /2$, we have
\begin{eqnarray}
\Delta \varphi = 4 \vert h \vert q \ e^{r} \sin \Omega t\,,
\label{enhancement}
\end{eqnarray}
and the signal is enhanced by a squeezing factor $ e^{r} $.
The enhancement can be traced back to the quantum nature of squeezed states, and we remind its connection to the enhancement of quantum noise due to squeezed GWs as predicted in \cite{Guerreiro2020, Parikh2021a, Parikh2021b} while predicting further that squeezing -- in particular squeezed-coherent states of GWs -- can lead to an enhancement in the \textit{signal} measurable by interferometric detectors.

If $ \phi_{0}  = 0 $ the signal becomes
\begin{eqnarray}
\Delta \varphi = 4  \vert h \vert q \ e^{-r} \  \sin \Omega t \,,
\label{supression}
\end{eqnarray}
and we find a suppression of the signal by a squeezing factor $ e^{-r} $. 
We note that the case of $ \phi_{0}  = 0 $ displays $ g^2_{GW} \approx e^{-2r} < 1 $ provided $ \vert h \vert^{2} \gg 8 e^{4r} \sinh^{2} r \cosh^{2} r $. 
For $ r = 1 $, $  g^2_{GW} \approx 1 / e^{2} \approx 0.13$, and $ \Delta \varphi / \Delta \varphi_{0} = 1 / e \approx 0.37$, indicating that a squeezed-coherent GW with suitable phase could display sub-Poissonian statistics -- and hence quantum behavior -- while only mildly attenuating the signal measured by LIGO. 

It is instructive to compare the enhanced and suppressed phases due to squeezed-coherent waves with the corresponding phase due to a coherent state GW.
Setting $ r = 0$ in \eqref{general_phase} the induced phase becomes \cite{Guerreiro2020},
\begin{eqnarray}
\Delta \varphi_{0} = 2 q \vert h \vert \sin \Omega t\,,
\label{coherent_phase_coupling}
\end{eqnarray}
and note the same phase was found in the precession of the Bloch vector in \eqref{Bloch_vec2}.
We can write the prefactor $ q \vert h \vert $ in terms of experimentally accessible quantities. A GW has energy-density given by $ E = (1/32\pi G) \Omega^{2} f^{2} $~\cite{Landau1975} where $ f $ denotes the GW strain for mode $\bm{k}$.
A single graviton has energy-density $ E_{g} = \Omega / V $, so a wave with energy density $ E $ has on average $ \langle N \rangle = E/ E_{g} $ gravitons. For coherent states the mean particle number is given by $ \vert h \vert = \sqrt{\langle N \rangle} $,  
so the acquired phase becomes
\begin{eqnarray}
\Delta \varphi_{0} =  \left( \dfrac{\omega}{4\Omega} \right)  \ f \sin \Omega t\,,
\label{coherent_phase}
\end{eqnarray}
in accordance to the classical prediction that a GW induces an optical phase proportional to its strain and oscillating at the GW frequency \cite{LIGO2017}. 

The phase \eqref{coherent_phase} allows us to connect our results with current gravitational wave detectors. Note the phase is proportional to the GW strain $ f $, which originates from writing the prefactor $ 2q\vert h \vert $ in \eqref{coherent_phase_coupling} in terms of experimentally accessible numbers. The same prefactor appears in the enhanced and suppressed phases \eqref{enhancement}, \eqref{supression} due to squeezed-coherent states, and in the limit that $ \vert h \vert^{2} \gg  \sinh^{2} r  $ we have $ \langle N \rangle \approx \vert h \vert^{2} $. Therefore, if the squeezed-coherent GWs have mean numbers of gravitons comparable to the coherent GWs currently detected by LIGO -- corresponding to strains of $f \approx 10^{-21}$ -- and only a modest squeezing parameter, say $ r \approx 1$, the signals due to these squeezed-coherent waves will be only slightly enhanced or suppressed with respect to signals currently detected.

Figure \ref{phases}a) shows the relative phase imprinted by a squeezed-coherent GW state for different values of the squeezing angle, in comparison to the signal produced by a coherent state GW. As shown above, for the particular values $ \phi_{0} = 0, \pi/2 $, the signal undergoes suppression and enhancement, respectively. For intermediate values of the squeezing angle (e.g. $ \phi_{0} = \pi/4 $) the signal acquires a characteristic form with asymmetric amplitude, which can also be used to single out the squeezed-coherent state from other possible quantum states of the GW. 
Next, in Figure \ref{phases}b) we show the maximum value of $ \Delta \varphi $ normalized by the corresponding maximum phase attained through interaction with a coherent GW, $ \mathrm{max} \left( \Delta \varphi_{0} \right) = \omega f / 4 \Omega $. As the squeezing parameter increases, the signal amplitude measured by the GW detector oscillates between enhancement and suppression according to the squeezing angle $ \phi_{0} $.

Detection of a signal as depicted in Figure \ref{phases} requires the coherent interaction between a GW and the optical modes on the order of a GW period. 
For GWs currently detected by LIGO, this corresponds to a photon lifetime of at least $ \sim \SI{300}{\mu s} $, amounting to a cavity decay rate smaller than $ \sim \SI{3}{kHz} $ but note that optical state reconstruction as described in the previous section only requires time derivatives of the optical density matrix elements near the initial time of the experiment, relaxing the need for high-Q cavities.

We revisit the qubit example discussed in the previous section, now for the case of a squeezed-coherent GW. With \eqref{coherent_overlap} and \eqref{kappa} we can calculate the Bloch vector of the optical field associated to a squeezed-coherent GW finding -- perhaps unsurprisingly at this point -- a combination of the coherent and squeezed results,
\begin{align}
\vec{r}_{\vert r, h \rangle}(t) &= \   e^{-2 e^{2r} q^{2} \sin^{2}\left(\Omega t \right)} \times \nonumber \\
& \left( \cos \left( \omega t + \Delta \varphi(t) \right),  \sin \left( \omega t + \Delta \varphi(t) \right), 0  \right) \, , \label{Bloch_vec4} 
\end{align}
where we have assumed $ \phi_{0} = \pi / 2 $ and large $ r $, for simplicity. We see the qubit inherits the shrinking factor associated to squeezing, plus an enhanced phase signal given in \eqref{enhancement}.

Similarly to the qubit model, an optical coherent state will acquire the phase signal manifest in the mean electric field $ \langle \mathcal{E}(t) \rangle$, and also noise characteristics manifest in 
electromagnetic (EM) field correlation functions such as $ g^{1}_{EM}(t) = \langle a^{\dagger}(0)a^{\dagger}(t) \rangle $, $ g^{2}_{EM} = \langle a^{\dagger}a^{\dagger} a a \rangle / \langle a^{\dagger} a \rangle^{2} $ and so on. Both signal and noise will provide information on the GW state. Calculation of the noise properties can be achieved through field theory methods, notably the Feynman-Vernon influence functional formalism \cite{Parikh2021b} or via the canonical quantization formalism. In certain situations Markovian approximations can be made; see for example \cite{Arani:2021oim} where the broadening of the power spectrum of a cavity due to inflationary squeezed states has been recently calculated.

As a final remark, we note similar effects are obtained if instead of a squeezed-coherent state we consider two-mode squeezed coherent states defined as 
\begin{eqnarray}
\vert \Psi \rangle_{12} = S_{12}(r,\phi(t)) D_{1}(h_{1}(t)) D_{2}(h_{2}(t)) \vert 0 \rangle\,, \nonumber \\
\label{two-mode state}
\end{eqnarray}
where $ S_{12}(r,\phi(t))  $ is the two-mode squeezing operator \cite{Schumaker1985} acting on modes `1' and `2' defined by wave-vectors $ \bm{k}_{1} $ and $ \bm{k}_{2} $, respectively, and $ D_{i}(h_{i}(t)) $ is a displacement operator acting on mode `$i$' responsible for generating a seed coherent state; see Appendix \ref{append:q-enahn-suppression}.
Interestingly, two-mode squeezed(-coherent) states have been previously studied in cosmology \cite{Albrecht:1992kf} and could have been generated in the early universe \cite{Kanno2019}.

\section{Conclusion}
\label{sec:conclusions}

In this work we have studied the optomechanical interaction of a GW in different quantum states -- vacuum, coherent, squeezed and squeezed-coherent -- with the quantized electromagnetic field in an optical cavity using the framework of effective field theory. 
We make no detailed assumption on the UV behavior of gravity, only requiring the existence of an energy range, well below the characteristic quantum gravity scale (assumed to be within a few order of magnitudes of the Planck scale) in which the use of effective theory is justified.

For the simplified situation in which the cavity state is given by a superposition of the vacuum and a single photon, referred to as an optical qubit, we have shown that trajectories of the associated Bloch vector are uniquely determined by the GW quantum state and are given by the combination of a shrinking factor associated to gravity-induced noise and precession motion resulting in a signal analogous to the one measured by conventional interferometric GW detectors.

Moreover, performing state tomography on an ensemble of electromagnetic quantum states which interacted with a single copy of a passing-by GW allows for the reconstruction of the GW quantum state, provided the state is Gaussian. This offers a unique experimental perspective on the interplay between quantum mechanics and gravity.

For the case in which the GW is in a squeezed-coherent state, the gravitational induced fluctuations can exhibit an exponential enhancement, as originally pointed out in \cite{Guerreiro2020, Parikh2021a, Parikh2021b}, and we show that the signal can also be exponentially enhanced or suppressed, depending on the squeezing phase. 
Squeezed-coherent GW states can also exhibit non-classical, sub-Poissonian graviton statistics. This opens the possibility of detecting quantum features associated to GWs in current or near-future detectors, provided possible sources of squeezed-coherent GWs emit with intensities comparable to the signals so far detected (strain $f \sim 10^{-21}$) and squeezing parameters of order one ($r \sim 1 $).

In quantum optics, squeezed-coherent states are produced when nonlinear mode-mode couplings or parametric modulations are present, which occur in dielectric crystals when the polarizability of the material is probed beyond linear order \cite{Boyd2008}.
In gravity, coupling between GW modes and between GWs and dynamical spacetime backgrounds naturally occurs due to the nonlinearity of General Relativity. Einstein's equations imply that the Riemann curvature tensor satisfies a nonlinear wave equation\footnote{The equation in components is $ R_{\alpha \beta \gamma \delta ; \mu}^{ \ \ \ \ \ \ \ \mu} = 2 R_{\alpha \beta \xi \mu} R_{\delta \ \ \gamma \ }^{ \ \ \xi \ \ \mu} + 2 R_{\alpha \xi \delta \mu} R_{\beta \ \ \gamma \ }^{ \ \ \xi \ \ \mu} - 2 R_{\alpha \xi \gamma \mu} R_{\beta \ \ \delta \ }^{ \ \ \xi \ \ \mu} $.}. Using the so-called Brill-Hartle average \cite{Brill1964}, one can split the curvature into a background and a GW part.
We can anticipate then at least two mechanisms for the parametric amplification of GWs: 
\begin{itemize}
    \item[I.] The covariant derivative in the wave operator contains connection coefficients which are dependent on the propagating GW perturbations, thus producing higher-order terms in the GW field. This can be related to the graviton-graviton scattering processes in the effective field theory approach to gravity.
    
    \item[II.] Even if quadratic couplings between the GWs are small, Einstein's equations contain terms coupling the curvature associated to GWs to the dynamical background curvature.
    If the radius of curvature of the background (given by the order of magnitude of the components of the Riemann curvature tensor) is comparable to the wavelength of the waves, such curvature-coupling may lead to parametric amplification of the GWs, which in a quantum mechanical description amounts to squeezing of the seed coherent state. One example of this is the generation of squeezed GWs in cosmological perturbation theory. 
\end{itemize} 

Examples of mechanism I. include high intensity GW astrophysical events \cite{Sawyer2020} and graviton-graviton scattering in field theory \cite{Grisaru1975}, while parametric processes of the type II. have been suggested in inflation \cite{PhysRevD.42.3413, PhysRevD.50.4807,PhysRevD.55.5917, Kanno2019} and the end stages of black hole evaporation \cite{natueHawking}. 

Noncoherent states of GWs are also expected to be produced via mechanism II. during the end stages of compact mergers \cite{Parikh2021a, Parikh2021b}.
Numerical simulations indicate that perturbed black holes can generate outgoing gravitational radiation with characteristic nonlinear features, notably mode-mode coupling, second harmonic generation, stronger-than-linear output and parametric instabilities \cite{Zlochower2003, Zimmerman2016}, all phenomena associated to squeezing of quantum noise \cite{Clerk2010}. 
In Kerr black holes, mode coupling can occur even at linear order \cite{Rostworowski2017} and more recently, a novel mechanism for the nonlinear generation of GWs during the ringdown phase of a binary black hole merger has been proposed \cite{Sberna2021}. 

Finally, quantum enhanced GW detectors, in the sense of \cite{Huang:2021pei}, might enable tests of the nonlinear features of gravity from the perspective of quantum mechanics.

\acknowledgments{We acknowledge Bruno Suassuna, Igor Brand\~ao, Bruno Melo, George Svetlichny and Leo Stein for helpful discussions. T.G. acknowledges the Coordenac\~ao de Aperfei\c{c}oamento de Pessoal de N\'ivel Superior - Brasil (CAPES) - Finance Code 001, Conselho Nacional de Desenvolvimento Cient\'ifico e Tecnol\'ogico (CNPq) and the FAPERJ Scholarship No. E-26/202.830/2019. J.R.W. acknowledges partial support from the LDRD programs of Lawrence Berkeley National Laboratory and Brookhaven National Laboratory, the U.S. Department of Energy, Office of Science, Office of Nuclear Physics, under contract number DE-AC02-05CH11231.  A.M.F. acknowledges  support from the ERC Advanced Grant GravBHs-692951, MEC grant FPA2016-76005-C2-2-P, and AGAUR grant 2017-SGR 754.}


\bibliographystyle{quantum}
\bibliography{main}

\onecolumn
\appendix

\section{Time evolution}
\label{append:time}
We review the time evolution operator here. The discrete-mode Hamiltonian (eqns. (2) and~(3)) in the main text can be exponentiated exactly. The result of exponentiating mode by mode is summarized in the unitary operator [See Eq. (2) in \cite{Brandao2020}]
\begin{eqnarray}
U_{\bm{k}} (t)  = e^{-i b^{\dagger}_{\bm{k}} b_{\bm{k}} \Omega t} e^{-i \omega a^{\dagger}a}  e^{iB_{k}(t) (a^{\dagger}a)^{2}} e^{q_{\bm{k}} a^{\dagger} a  ( \gamma_{k} b_{k} - \gamma_{k}^{*} b^{\dagger}_{k}   )}
\label{complete_Uk}
\end{eqnarray}
where $ \gamma_{k} = (1 - e^{-i\Omega_{k} t}) $ and $B_{k}(t) = q_{\bm{k}}^{2} \left(   \Omega_{k} t - \sin \Omega_{k} t  \right)   $. From now on we will work on a single-mode and omit the mode subscript $ \bm{k} $.

A useful formula is 
\begin{eqnarray}
e^{-\lambda a^{\dagger} a} a e^{\lambda a^{\dagger} a} = a e^{\lambda}
\end{eqnarray}
We can ignore the free electromagnetic evolution in \eqref{complete_Uk} as it only generates a linear time-dependent phase on the electromagnetic annihilation operator. 
As in the main text, we also ignore the nonlinear  $B_{k}(t) $ term, as it has a negligible effect scaling quadratically and higher in $ q_{\bm{k}} $. The evolution of a GW state under \eqref{complete_Uk} then reads
\begin{align}
U_{\bm{k}} (t)  \vert \Psi \rangle  &=  e^{-i b^{\dagger} b \Omega t} e^{q a^{\dagger} a  ( \gamma b - \gamma^{*} b^{\dagger}  )} e^{i b^{\dagger} b \Omega t} e^{-i b^{\dagger} b \Omega t} \vert \Psi \rangle \nonumber \\
&= e^{q a^{\dagger} a  ( \gamma e^{i\Omega t} b - \gamma^{*} e^{-i\Omega t} b^{\dagger}  )} e^{-i b^{\dagger} b \Omega t} \vert \Psi \rangle \nonumber \\
&=  e^{q a^{\dagger} a  ( \gamma b^{\dagger} - \gamma^{*}  b  )}  \vert \Psi(t) \rangle
\label{time_evolution2}
\end{align}
where we have defined the time-evolving state $ \vert \Psi(t) \rangle = e^{-i b^{\dagger} b \Omega t} \vert \Psi \rangle $.
We can regard \eqref{time_evolution2} as an operator acting on the photon annihilation operator by conjugation, which results in 
\begin{eqnarray}
a(t) = \langle \Psi \vert U_{\bm{k}} (t)^{\dagger} a U_{\bm{k}} (t) \vert \Psi \rangle = a \langle \Psi(t) \vert D(q\gamma) \vert \Psi(t) \rangle
\end{eqnarray}
For the case of a squeezed-coherent state, for example,
\begin{eqnarray}
\vert \Psi \rangle = S(r,\phi_{0}) D(\vert h \vert) \vert 0 \rangle
\end{eqnarray}
from which we can calculate
\begin{eqnarray}
\vert \Psi(t) \rangle &=&  e^{-i b^{\dagger} b \Omega t} \vert \Psi \rangle \nonumber \\
&=&  e^{-i b^{\dagger} b \Omega t}     S(r,\phi_{0}) D(\vert h \vert)   e^{i b^{\dagger} b \Omega t}   e^{-i b^{\dagger} b \Omega t} \vert 0 \rangle \nonumber \\
&=& e^{-i b^{\dagger} b \Omega t}     S(r,\phi_{0}) D(\vert h \vert)   e^{i b^{\dagger} b \Omega t}   \vert 0 \rangle \nonumber \\
&=& S(r,\phi) D( h ) \vert 0 \rangle
\end{eqnarray}
where $ \phi = \phi_{0} - \Omega t $ and $ h = \vert h \vert e^{-i\Omega t} $, as considered in the main text. 

\section{Multi-mode corrections}
\label{multi-mode-corrections}

In the main text, we consider the effects of a single-mode GW in a state with non-vanishing mean number of gravitons interacting with the detector and neglect the effects of the additional modes in the vacuum state. For completeness, we briefly review the results from \cite{Guerreiro2020} which justify this single-mode approximation. 

The time evolution of the coupled optical-GW system is given by a product of the operators $ U_{k}(t) $, defined in \eqref{Uk} in the main text. 
We now wish to estimate the effect of vacuum modes upon electromagnetic operators. As an example, consider the optical annihilation operator $ a $.
Its time evolution is then given by
\begin{eqnarray}
a(t) &=& \left( \prod_{k} U_{k}(t)   \right)^{\dagger} a \left( \prod_{k} U_{k}(t)   \right) \nonumber \\
&=& \prod_{k} e^{iB_{k}(t) a^{\dagger}a} e^{i B_{k}(t)/2} e^{q_{k}a^{\dagger}a\left( \gamma_{k}b^{\dagger} - \gamma_{k}^{*} b_{k}   \right)} a  \nonumber \\
&=& \prod_{k} e^{iB_{k}(t) a^{\dagger}a} e^{i B_{k}(t)/2} D(q_{k}\gamma_{k}) a
\label{annihilation_op}
\end{eqnarray}
where $ D(q_{k}\gamma_{k}) $ is the GW displacement operator. Consider now the GW field to be in the vacuum state
\begin{eqnarray}
\vert 0 \rangle = \prod_{k} \vert 0_{k} \rangle 
\end{eqnarray}
where $ \vert 0_{k} \rangle  $ denotes the vacuum state in mode $ k $. The annihilation operator of an optical mode interacting with such gravitational vacuum can then be written as 
\begin{eqnarray}
a(t) = e^{i \mathcal{F}(t) a^{\dagger} a} e^{i\mathcal{F}(t)/2} \mathcal{G}(t)a
\end{eqnarray}
where 
\begin{eqnarray}
\mathcal{F}(t) = \sum_{k} B_{k}(t)
\end{eqnarray}
and 
\begin{eqnarray}
 \mathcal{G}(t) = \prod_{k} \langle 0_{k} \vert D(q_{k}\gamma_{k}) \vert 0_{k} \rangle = \exp \left[ -\dfrac{1}{2} \sum_{k} q_{k}^{2}\vert \gamma_{k}\vert^{2} \right]
\end{eqnarray}
Physically, the expressions above must be cut-off at a maximum and minimum frequencies $ \Omega_{k} = \vert k \vert $ to which the detector can be sensitive in principle \cite{Parikh2021b}; for illustration purposes, we will exaggerate these infrared and ultraviolet cut-offs to the Hubble and Planck energies, $ E_{\rm IR} $ and $ E_{\rm pl}$, respectively. Recovering the continuous limit yields, up to numerical factors of order one,
\begin{eqnarray}
\mathcal{F}(t) &=& \int \dfrac{d^{3} \bm{k}}{\sqrt{(2\pi)^{3}}} 2\omega^{2} \left( \dfrac{8\pi G}{k^{3}} \right) \left( \Omega_{k} t - \sin \Omega_{k} t  \right) \nonumber \\
& \approx & \left( \dfrac{\omega}{E_{\rm pl}}   \right)^{2} \left( \int_{E_{\rm IR}}^{E_{\rm pl}} dk   \right) t \approx \left( \dfrac{\omega}{E_{\rm pl}}   \right) \omega t \label{phase_noise}
\end{eqnarray}
where we have used $ E_{\rm pl} \gg E_{\rm IR} $, considered large times $ t \gg E_{\rm IR}^{-1} $ and neglected the bounded term $ \sin \Omega_{k} t  $.
Corrections to the optical annihilation operator due to the phase $ \mathcal{F}(t) $ are then on the order of $ (\omega / E_{\rm pl}) $ and only become relevant for optical energies close to the Planck energy.
Similarly,
\begin{eqnarray}
\mathcal{G}(t) &=& \exp\left(  -\dfrac{1}{2} \int  \dfrac{d^{3} \bm{k}}{\sqrt{(2\pi)^{3}}} \omega^{2} \left( \dfrac{8\pi G}{k^{3}} \right) \vert \gamma_{k} \vert^{2} \right) \nonumber \\
& \approx & \exp \left(  - 2 \left( \dfrac{\omega}{E_{\rm pl}}   \right)^{2}  \int_{E_{\rm IR}}^{E_{\rm pl}} \dfrac{dk}{k}  \right) \nonumber \\
&=& \exp \left(  - 2\left( \dfrac{\omega}{E_{\rm pl}}   \right)^{2} \ln \left( \dfrac{E_{\rm pl}}{E_{\rm IR}} \right) \right)
\end{eqnarray}
where we have considered the worst-case approximation $ \vert \gamma_{k} \vert^{2} = 2 \left( 1 - \cos \Omega_{k} t \right) \sim 4 $. Notice that the ratio of ultraviolet to infrared cut-offs is $ E_{\rm pl} / E_{\rm IR} \approx 10^{62} $, giving a correction to $ a(t) $ approximately proportional to $ (\omega / E_{\rm pl})^{2} $, a tremendously small number! 

Instead of the vacuum, we could consider all modes to be populated by a thermal state at about \SI{1}{K}, the expected temperature for the cosmic GW background \cite{Allen1993}. This would not alter $ \mathcal{F}(t) $, which is state-independent, so the estimates in \eqref{phase_noise} remain. The $ \mathcal{G}(t) $ term, however, would acquire a correction factor at most $ e^{-q_{\rm pl}^{2} (2 \bar{n} + 1)} $, with $ \bar{n} = 1 / (e^{\hbar \Omega_{k} / k_{B}T} - 1) $ and $ q_{\rm pl}
$ the GW coupling strength at the Planck frequency. \cite{Guerreiro2020}. For a GW mode of \SI{10}{Hz}, peak of the expected cosmic GW background spectrum, this correction factor amounts to $ \approx e^{-10^{-47}} $. 

All in all, these estimates show that the effect of modes which are not populated by states with a very large mean number of gravitons upon optical observables is completely negligible. In other words, decoherence due to the gravitational vacuum, or even due to the thermal background of GWs is very weak, which is consistent with previous results \cite{Dyson2013, Blencowe2013}.

\section{Gravitational decoherence of cavity modes}
\label{appendix-2-decoherence}

Here we explore the idea that terms quadratic in the dimensionless coupling $ q $ appearing in the exponent of (10) are associated to GW-induced decoherence.
Let us consider the electromagnetic-gravitational wave (EM-GW) system prepared at time $t=0$ in the state
\begin{equation}
    \ket{\Psi\left(0\right)} = \frac{\ket{0}_\text{EM}+\ket{N}_\text{EM}}{\sqrt{2}} \otimes \ket{0}_\text{GW}
\end{equation}
This provides a simplified model of a qubit (defined by the span of the $\lbrace \vert 0 \rangle , \vert N \rangle \rbrace$ subspace) exhibiting features similar to the electromagnetic coherent states considered in the main, but with easily calculable coherence terms.

The initial state $ \ket{\Psi\left(0\right)} $ can be evolved directly using operator~(7) resulting in the state at arbitrary times,
\begin{equation}
    \ket{\Psi\left(t\right)} = \frac{\ket{0}\ket{0}+\ket{N}\ket{qN\gamma}}{\sqrt{2}} 
\end{equation}
where we omit the subscripts by intending that the left (right) kets refer to EM (GW), respectively; the order reverses when considering bras instead of kets. 
Notice that the time dependence is contained in the function $\gamma\left(t\right)$, and moreover $\ket{qN\gamma}$ is a coherent state. The total density matrix associated to $ \ket{\Psi\left(t\right)} $ reads,
\begin{align}
    \rho\left(t\right) = \frac{1}{2}\left(
      \ket{0}\ket{0}        \bra{0}\bra{0} 
    + \ket{0}\ket{0}        \bra{qN\gamma}\bra{N}  
    +  \ket{N}\ket{qN\gamma} \bra{0}\bra{0}
    + \ket{N}\ket{qN\gamma} \bra{qN\gamma}\bra{N}
    \right)
     \ ,
\end{align}
from which we can trace out the GW subsystem to obtain the density matrix associated to the electromagnetic field,
\begin{equation}\label{eq:densityMatrixTracedOut}
    \rho_\text{EM}\left(t\right) = \begin{pmatrix}
    \frac{1}{2} & \rho_{01} \\
    \rho_{01}^{*} & \frac{1}{2}
    \end{pmatrix}
\end{equation}
where the coherence term $\rho_{01}=\braket{0}{qN\gamma} = \exp{-\frac{1}{2}q^2N^2\left|\gamma\right|^2}$ oscillates with an amplitude proportional to $q^2$. This supports the idea that exponents proportional to the dimensionless coupling squared are associated to GW-induced decoherence.

To further explore this idea, let us now consider the GW field in a coherent state,
\begin{equation}
    \ket{\Psi\left(0\right)} = \frac{\ket{0} +\ket{N}}{\sqrt{2}} \otimes \ket{\alpha}
\end{equation}
Time evolution yields,
\begin{equation}
    \ket{\Psi\left(t\right)} = \frac{\ket{0}\ket{\alpha}+\ket{N}D\left(qN\gamma\right)\ket{\alpha}}{\sqrt{2}}
\end{equation}
Furthermore,
\begin{eqnarray}
    D\left(qN\gamma\right)\ket{\alpha} &=& D\left(qN\gamma\right)D\left(\alpha\right)\ket{0} \\
    &=& e^{\frac{1}{2}Nq\left(\gamma\alpha^*-\gamma^*\alpha\right)}\ket{qN\gamma+\alpha}
\end{eqnarray}
and the density matrix now becomes,
\begin{equation}
\begin{split}
    \rho\left(t\right) = \text{ } &\frac{1}{2}\ket{0}\ket{\alpha}\bra{\alpha}\bra{0} \\
    &+ \frac{1}{2} e^{-\frac{1}{2}Nq\left(\gamma\alpha^*-\gamma^*\alpha\right)} \ket{0}\ket{\alpha} \bra{qN\gamma+\alpha}\bra{N} \\
    &+ \frac{1}{2} e^{\frac{1}{2}Nq\left(\gamma\alpha^*-\gamma^*\alpha\right)} \ket{N}\ket{qN\gamma+\alpha} \bra{\alpha}\bra{0} \\
    &+ \frac{1}{2}\ket{N}\ket{qN\gamma+\alpha} \bra{qN\gamma+\alpha}\bra{N}
    \end{split}
\end{equation}
Tracing out the GW states as before yields the same form as~\eqref{eq:densityMatrixTracedOut} with $\rho_{01}$ given by
\begin{equation}
    \rho_{01} = e^{-\frac{1}{2}\left[q^2N^2\left|\gamma\right|^2+qN\left(\gamma^*\alpha-\gamma\alpha^*\right)\right]}
    \label{coherent_coherence}
\end{equation}
We see the interaction between the EM field and a a quantized single-mode GW is given by the $ q^{2} $ term present when the GW-field is in the vacuum state plus an oscillating factor linear in $ q $ and proportional to the GW amplitude.

Equation \eqref{coherent_coherence} naturally allows us to evaluate the expression of $\rho_{01}$ for a single-mode GW in a thermal state with mean number of gravitons $\overline{n}$,
\begin{eqnarray}
    \rho_{01} &=& e^{-\frac{1}{2}q^2N^2\left|\gamma\right|^2} \int \frac{d^2\alpha}{\pi \overline{n}} e^{-\frac{1}{2}\left[\frac{\left|\alpha\right|^2}{\overline{n}}+qN\left(\gamma^*\alpha-\gamma\alpha^*\right)\right]} \nonumber \\
    &=& e^{-\frac{1}{2}q^2N^2\left|\gamma\right|^2 \left(1+\overline{n}\right)}
    \label{final_coherence}
\end{eqnarray}
This expression is valid for a single-mode thermal state. We could generalize it to a continuum of GW modes in thermal states, with the initial state 
\begin{eqnarray}
\rho(0) = \vert \psi_{0} \rangle \langle \psi_{0} \vert \otimes \prod_{\bm{k}} \rho^{\text{Th}}_{\bm{k}}
\end{eqnarray}
where $ \vert \psi_{0} \rangle = \frac{\ket{0}+\ket{N}}{\sqrt{2}}$ and $\rho^{\text{Th}}_{\bm{k}} $ is the thermal density matrix for mode $\bm{k}$.
For such state, \eqref{final_coherence} generalizes to 
\begin{eqnarray}
&\ & \prod_{\bm{k}}  e^{-\frac{1}{2}q_{\bm{k}}^2N^2\left|\gamma_{\bm{k}}\right|^2\left(1+\overline{n}_{\bm{k}}\right)} =\exp \left( -\frac{1}{2}N^2 \sum_{\bm{k}} q_{\bm{k}}^2\left|\gamma_{\bm{k}}\right|^2\left(1+\overline{n}_{\bm{k}}\right)     \right)
\end{eqnarray}
If the GW thermal states are assumed to be highly populated $ \overline{n}_{\bm{k}} \gg 1 $ ($ T \rightarrow \infty $), we can write $ \left(1+\overline{n}_{\bm{k}}\right) \approx K_{B} T / \Omega_k $. Substituting $ q_{\bm{k}} = (\omega / \Omega_k) \sqrt{8\pi G / \Omega_k V} / 4 $, $ \vert \gamma_{\bm{k}} \vert^{2} = 4 \sin^{2}\left( \frac{\Omega_k t}{2} \right) $ and employing the density of modes for a bosonic field in a volume $ V $ \cite{Scully1997} given by $ \mathcal{D}(\Omega) \propto V \Omega^{2} d\Omega $ we find, in the continuum limit and after summing over all modes,
\begin{eqnarray}
\rho_{01} \approx \exp \left\{  k_{B}T \left( \dfrac{N\omega}{E_{P}}\right)^{2} \zeta \int d\Omega \dfrac{\sin^{2}\left( \frac{\Omega t}{2} \right)}{\Omega^{2}} \right\}
\label{coherence_multimode}
\end{eqnarray}
where $ E_{\rm pl} = 1 / \sqrt{G} $ is Planck's energy, $\zeta$ is a factor of order one and we note that the integral in the exponent is the same as Eq. (22) in \cite{Eckert1996} and Eq. (23) in \cite{Vedral2020_decoherence}. 
From the scaling of the integral in \eqref{coherence_multimode} we see that
\begin{eqnarray}
\rho_{01} \approx e^{- \Gamma t}
\end{eqnarray}
where 
\begin{eqnarray}
\Gamma \propto k_{B}T \left( \dfrac{\Delta E}{E_{\rm pl}}\right)^{2}
\end{eqnarray}
up to a factor of order one, where we have defined the energy difference $ \Delta E = N\omega $. With this we recover the scaling of the decoherence of a superposition with energy difference $ \Delta E $ in the presence of a thermal GW background with temperature $ T $ as obtained in \cite{Blencowe2013}.

\section{Reconstruction of Gaussian GW states from optical modes}
\label{append:reconstruction}

As explained in the main text, the reduced density matrix elements of an optical state which interacted with a single mode quantum GW state are proportional to 
\begin{align}
\mathcal{I}_{n}(t) = \langle \Psi \vert D(-qn\gamma^{*}) \vert \Psi \rangle =   \mathrm{Tr}\left(  \sigma  D(-nq\gamma^{*}) \right) =  \mathrm{Tr}\left(  \sigma  e^{-qn\left(  \gamma^{*} b^{\dagger} - \gamma b  \right)} \right)
\label{I_quantity}
\end{align}
These correspond to the quantum characteristic function associated to the GW evaluated at certain contours in the complex plane, and as such contain information on the GW quantum state. By measuring different components of the optical reduced density matrix, we can then obtain such information and partially reconstruct the GW quantum state. As we now show, if the state is Gaussian reconstruction is perfect. 

The idea behind the GW state reconstruction is to obtain relations between GW field correlators and time derivatives of $ \mathcal{I}_{n}(t) $ evaluated at $ t = 0 $. Inverting these relations, we can obtain the first and second moments of the gravitational field quadratures. 
To see that, we Taylor expand $ \mathcal{I}_{n}(t) $ in time using,
\begin{align}
e^{n A (e^{it} - 1)} = 1 + i n A t - \dfrac{1}{2}\left( nA \left( nA + 1 \right) \right) t^{2} + \nonumber \\
 - \dfrac{i}{6} n A \left(   n^{2} A^{2} + 3 n A + 1  \right) t^{3} + \mathcal{O}(t^{4}) \\
e^{-n B (e^{-it} - 1)} = 1 + i n B t - \dfrac{1}{2}\left( nB \left( nB - 1 \right) \right) t^{2} + \nonumber \\
 - \dfrac{i}{6} n B \left(   n^{2} B^{2} - 3 n B + 1  \right) t^{3} + \mathcal{O}(t^{4}) 
\end{align}
Observe also that 
\begin{align}
\mathcal{I}_{n}(t) = e^{-q^{2}n^{2}\vert \gamma \vert^{2}/2}  C(-qn\gamma^{*}, -qn\gamma)  \nonumber \\
= \left(  1 - \frac{1}{2}q^{2}n^{2}t^{2} + \mathcal{O}(q^{4}) \right) C(-qn\gamma^{*}, -qn\gamma)
\label{density_element}
\end{align}
We now proceed order-by-order in $ t $.
\\

\textbf{$ t $-terms:} The first order derivative of \eqref{density_element} gives us
\begin{eqnarray}
\alpha_{1}(n) =  i n q \ \langle X \rangle
\end{eqnarray}
From which we obtain
\begin{eqnarray}
\langle qX \rangle = - i \alpha_{1}(1)
\end{eqnarray}

\textbf{$ t^{2} $-terms:} The second order derivative of \eqref{density_element} yields,
\begin{eqnarray}
\alpha_{2}(n) = - n^{2} u_{1} - n u_{2}  - n^{2}q^{2}
\end{eqnarray}
where 
\begin{eqnarray}
u_{1} &=&  q^{2} \langle b^{\dagger 2} + b^{2} + 2 b^{\dagger} b   \rangle   \\ 
u_{2} &=& - q \langle  b - b^{\dagger}  \rangle
\end{eqnarray}
We can consider different values of $ n $, say $ n = 1, 2 $ to obtain two independent relations which can be used to solve for $ u_{1}, u_{2} $. We find,
\begin{eqnarray}
u_{1} &=& \alpha_{2}(1) - \frac{1}{2} \alpha_{2}(2) - q^{2} \\
u_{2}  &=& -2\alpha_{2}(1) + \frac{1}{2} \alpha_{2}(2) 
\end{eqnarray}
Now, observe that
\begin{eqnarray}
u_{1} = q^{2} \langle b^{\dagger 2} + b^{2} + 2 b^{\dagger} b   \rangle = \langle q^{2}X^{2} \rangle - \frac{i}{2} \langle \left[ qX,qY  \right] \rangle
\end{eqnarray}
We note $ \left[ b, b^{\dagger} \right] = \frac{i}{2} \left[ X, Y \right] = \mathds{1} $, but as we will see, it is sometimes convenient to keep expressions in terms of the commutator when taking the continuous limit.

We find,
\begin{eqnarray}
\langle q^{2}X^{2} \rangle = \alpha_{2}(1) - \frac{1}{2} \alpha_{2}(2)
\label{X2}
\end{eqnarray}

Similarly, 
\begin{eqnarray}
u_{2} =  -q \langle  b - b^{\dagger}  \rangle = i\langle qY \rangle =  \frac{1}{2} \alpha_{2}(2) - 2\alpha_{2}(1)
\end{eqnarray}

\textbf{$ t^{3} $-terms:} At third order we obtain 
\begin{eqnarray}
- \alpha_{3}(n) = 6n^{3} v_{1} + 6n^{2} v_{2} - n \alpha_{1}(1) 
\end{eqnarray}
where
\begin{eqnarray}
v_{1} &=& \frac{iq^{3}}{6} \langle b^{3} + 3b^{\dagger}b^{2} + 3b^{\dagger 2}b + b^{\dagger 3} + 3\left( b + b^{\dagger}  \right) \rangle \\
v_{2} &=& \frac{iq^{2}}{2}\langle b^{\dagger 2} - b^{2} \rangle
\end{eqnarray}
and once again, using different values of $ n $ we can recover $ v_{1}, v_{2} $. For Gaussian states, it suffices to determine $ v_{2} $. We have,
\begin{eqnarray}
v_{2} = - \frac{\alpha_{3}(1)}{3} + \frac{\alpha_{3}(2)}{24} - \frac{\alpha_{1}(1)}{4}
\end{eqnarray}
Using the relation
\begin{eqnarray}
\langle b^{\dagger 2} - b^{2} \rangle = i \langle XY \rangle - \frac{i}{2} \langle \left[ X, Y \right] \rangle
\end{eqnarray}
we find,
\begin{align}
\langle qX qY \rangle - \frac{1}{2}\langle \left[ qX, qY \right] \rangle = \frac{2}{3} \alpha_{3}(1) - \frac{1}{12} \alpha_{3}(2) + \frac{1}{2}\alpha_{1}(1)
\end{align}

We can go on to higher orders, and determine more normal-ordered expectation values of graviton creation and annihilation operators, but for Gaussian quantum states, determining the moments $ \langle X \rangle, \langle Y \rangle , \langle X^{2} \rangle $ and  $ \langle XY \rangle $ are sufficient; we still need the expecation value $ \langle Y^{2} \rangle $, which can be obtained either by normalization of the Wigner function or by additional measurements of the mean number of gravitons $ \langle N \rangle = \langle b^{\dagger} b \rangle $, since $ \langle Y^{2} \rangle = 2 + 4\langle b^{\dagger}b \rangle - \langle X^{2} \rangle $. This leads to the first and second moments discussed in the main text.

\section{Quantum enhancement and suppression of GW signals}
\label{append:q-enahn-suppression}

We provide further details on the calculations presented in the main text. We follow the conventions used in \cite{Schumaker1985}. Useful formulae are:
\begin{eqnarray}\label{eq:usefulFormula}
\langle \gamma  \vert D(\alpha) \vert \beta  \rangle = e^{-\vert \alpha \vert^{2}/2} e^{-\vert \beta \vert^{2}/2} e^{-\vert \gamma \vert^{2}/2} e^{\alpha \gamma^{*} - \alpha^{*} \beta + \beta \gamma^{*}}
\label{overlap}
\end{eqnarray}
where $ D(\alpha) $ is a displacement operator. Products of displacement operators satisfy:
\begin{eqnarray}
D^{\dagger}(\beta)D(\alpha) &=& D(-\beta) D(\alpha)   = e^{\frac{(\alpha\beta^{*} - \beta\alpha^{*})}{2}} D(\alpha - \beta)
\end{eqnarray}

Moreover,
\begin{eqnarray}
S(r,\phi) = \exp \left( \dfrac{1}{2} r \left(   a^{2} e^{-2i\phi} - a^{\dagger 2} e^{2i\phi}  \right)    \right)
\end{eqnarray}
is the single mode squeezing operator with squeezing parameter $ r $ and angle $ \phi $, and
\begin{eqnarray}
e^{i\theta a^{\dagger} a} S(r,\phi) e^{-i\theta a^{\dagger} a} = S(r,\phi + \theta)
\end{eqnarray}
gives us the time evolution of the squeezing parameter in the Heisenberg picture. Another useful formula is
\begin{eqnarray}
S^{\dagger}(r,\phi) D(\alpha) S(r,\phi) = D(\alpha \cosh r + \alpha^{*} e^{2i\phi} \sinh r)
\label{composition}
\end{eqnarray}

Define two modes `1' and `2'. The above formulas generalize to the case of a two-mode squeezing operator,
\begin{eqnarray}
S_{12}(r,\phi) =  \exp \left( \dfrac{1}{2} r \left(   b_{1}b_{2} e^{-2i\phi} - b_{1}^{\dagger }b_{2}^{\dagger } e^{2i\phi}  \right)    \right)
\end{eqnarray}
For instance, we have
\begin{eqnarray}
S_{12}^{\dagger} \  b_{1} \  S_{12} &=& b_{1} \cosh r + b_{2}^{\dagger} e^{2i\phi} \sinh r \\
S_{12}^{\dagger} \ b_{2} \  S_{12} &=& b_{2} \cosh r + b_{1}^{\dagger} e^{2i\phi} \sinh r 
\end{eqnarray}
which leads to the generalization of \eqref{composition},
\begin{align}
 S_{12}^{\dagger} & D_{1}(\alpha) D_{2}(\beta) S_{12} =  D_{1}(\alpha \cosh r) D_{1}(-\beta^{*}e^{2i\phi} \sinh r)  D_{2}(\alpha^{*} e^{2i\phi} \sinh r) D_{2}(\beta \cosh r)
\end{align}

We are interested in calculating the time-development of a single mode cavity electric field under the influence of a quantum GW. Here we focus on the case of a two-mode squeezed-coherent GW, since it generalizes the one mode calculation. 
Considering terms which generate only linear signals in the gravitational coupling $ q_{k} $, the time evolution operator for two GW modes `1' and `2' reads
\begin{eqnarray}
U(t)  \approx e^{q_{1} a^{\dagger} a  ( \gamma_{1} b^{\dagger}_{1}  - \gamma_{1}^{*} b_{1} )} e^{q_{2} a^{\dagger} a  ( \gamma_{1} b^{\dagger}_{2} - \gamma_{1}^{*} b_{2}   )}
\end{eqnarray}
which for an uncorrelated electromagnetic-gravitational state leads to the electric field
\begin{eqnarray}
\mathcal{E}(t) = 
\sqrt{\dfrac{\omega}{V_{c}}} \left(    \dfrac{ _{12}\langle \Psi(t) \vert D_{1}(q_{1}\gamma_{1}) D_{2}(q_{2}\gamma_{2}) \vert \Psi(t) \rangle_{12} a + h.c.}{\sqrt{2}} \right)
\label{field}
\end{eqnarray}
The two-mode GW induced phase originates from the term 
\begin{equation}
 _{12}\langle \Psi(t) \vert D_{1}(q_{1}\gamma_{1}) D_{2}(q_{2}\gamma_{2}) \vert \Psi(t) \rangle_{12} 
\end{equation}
and thus we focus on calculating it explicitly. 

We shall consider the two-mode squeezed-coherent state defined as in the main text,
\begin{eqnarray}
\vert \Psi(t) \rangle_{12} = S_{12}(r,\phi(t)) D_{1}(h(t)) \vert 0 \rangle
\end{eqnarray}
where $ \phi(t) = \phi_{0} - \Omega t $ and $ h(t) = \vert h \vert e^{-i\Omega t} $. 
In cosmology applications \cite{Albrecht1994}, we are interested in modes `1' and `2' defined by the wave-vectors $ -\bm{k} $ and $ \bm{k} $, respectively. Hence we have $ q_{1} = q_{2} = q $ and $ \gamma_{1} = \gamma_{2} = \gamma $. Therefore,
\begin{eqnarray}
& \ & _{12}\langle \Psi(t) \vert D_{1}(q\gamma) D_{2}(q\gamma) \vert \Psi(t) \rangle_{12} = \\ \nonumber
\\ 
&=&  _{2}\langle 0 \vert _{1}\langle h(t) \vert S_{12}^{\dagger} D_{1}(q\gamma) D_{2}(q\gamma) S_{12} \vert h(t) \rangle_{1} \vert 0 \rangle_{2} \nonumber \\ \nonumber
\\
&=& \underbrace{\langle h(t) \vert D_{1}(q\gamma \cosh r)D_{1}(-q\gamma^{*}e^{2i\phi}\sinh r)  \vert h(t) \rangle}_{\text{(I)}} \nonumber \\
&& \times  \ \underbrace{\langle 0 \vert D_{2}(q\gamma^{*}e^{2i\phi}\sinh r) D_{2}(q\gamma \cosh r) \vert 0 \rangle}_{\text{(II)}} \nonumber
\label{I}
\end{eqnarray}
Defining the coherent state $ \vert q\gamma \cosh r \rangle = D(q\gamma \cosh r)\vert 0 \rangle $ we can rewrite (II) as,
\begin{eqnarray}
\langle 0 \vert D(q\gamma^{*} e^{2i\phi} \sinh r) \vert q\gamma \cosh r \rangle
\end{eqnarray}
and using \eqref{overlap} we see that it only yields terms of order $ \mathcal{O}(e^{-q^{2}}) $. To leading order in $ q $, then, (II) can be approximated as unity.
As for the term (I) in \eqref{I} we write
\begin{eqnarray}
& & \langle  h  \vert  D(q\gamma \cosh r)D(-q\gamma^{*}e^{2i\phi} \sinh r)\vert h \rangle = \nonumber \\ \nonumber
\\
&=& \langle h \vert D^{\dagger}(-q\gamma \cosh r) D(-q\gamma^{*}e^{2i\phi} \sinh r) \vert h \rangle \nonumber \\ \nonumber
\\
&=& \mathcal{O}(e^{-q^{2}}) \times  \langle h \vert D\big( q (\gamma \cosh r - \gamma^{*} e^{2i\phi} \sinh r )   \big)  \vert h \rangle \nonumber \\ \nonumber
\\
&\approx & \langle h \vert D\big( q (\gamma \cosh r - \gamma^{*} e^{2i\phi} \sinh r )   \big)  \vert h \rangle
\end{eqnarray}
This is equivalent, up to a phase, to Eq. \eqref{coherent_overlap} in the main text.

\end{document}